\documentclass[11pt, letter]{article}

\usepackage{amsmath}
\usepackage{amssymb}
\usepackage{graphicx}

\def\>{\ensuremath{\rangle}}
\def\<{\ensuremath{\langle}}
\def\ra{\ensuremath{\rightarrow}}

\newtheorem{thm}{Theorem}[section]
\newtheorem{cor}{Corollary}[section]
\newtheorem{lem}{Lemma}[section]

\newtheorem{defn}{Definition}[section]
\newtheorem{prop}{Proposition}[section]
\newtheorem{exam}{Example}[section]

\begin{document}

\title{Quantum Loop Programs\thanks{This work was partly supported by FCT and EU FEDER POCTI
(via QuantLog POCI/MAT/55796/2004 Project), the National Natural
Science Foundation of China (Grant No: 60321002, 60496321) and the
Key Grant Project of Chinese Ministry of Education (Grant No:
10403).}}

\author{Mingsheng Ying\thanks{Email: yingmsh@tsinghua.edu.cn.}, and Yuan Feng\thanks{Email:
feng-y@tsinghua.edu.cn.}\\
\\  \small{State Key Laboratory of Intelligent Technology and
Systems,}\\
\small{ Department of Computer Science and Technology, }\\
\small{ Tsinghua University, Beijing, China }}
\date{}
 \maketitle

\begin{abstract}
Loop is a powerful program construct in classical computation, but
its power is still not exploited fully in quantum computation. The
exploitation of such power definitely requires a deep understanding
of the mechanism of quantum loop programs. In this paper, we
introduce a general scheme of quantum loops and describe its
computational process. The notions of termination and almost
termination are proposed for quantum loops, and the function
computed by a quantum loop is defined. To show their expressive
power, quantum loops are applied in describing quantum walks.
Necessary and sufficient conditions for termination and almost
termination of a general quantum loop on any mixed input state are
presented. A quantum loop is said to be (almost) terminating if it
(almost) terminates on any input state. We show that a quantum loop
is almost terminating if and only if it is uniformly almost
terminating. It is observed that a small disturbance either on the
unitary transformation in the loop body or on the measurement in the
loop guard can make any quantum loop (almost) terminating. Moreover,
a representation of the function computed by a quantum loop is given
in terms of finite summations of matrices. To illustrate the notions
and results obtained in this paper, two simplest classes of quantum
loop programs, one qubit quantum loops, and two qubit quantum loops
defined by controlled gates, are carefully examined.
\end{abstract}

\section{Introduction}

One of the most striking advances in quantum computing was made by
Shor \cite{Sh94} in 1994. By exploring the power of quantum
parallelism, he discovered a polynomial-time algorithm on quantum
computers for prime factorization of which the best known algorithm
on classical computers is exponential. In 1996, Grover \cite{Gr96}
offered another apt killer of quantum computation, and he found a
quantum algorithm for searching a single item in an unsorted
database in square root of the time it would take on a classical
computer. Since both prime factorization and database search are
central problems in computer science and the quantum algorithms for
them are highly faster than the classical ones, Shor and Grover's
discoveries indicated that quantum computation offers a way to
accomplish certain computational tasks much more efficiently than
classical computation and thus stimulated an intensive investigation
on quantum computation. After that, quantum computation has been an
extremely exciting and rapidly growing field of research. In
particular, a substantial effort has been made to find new quantum
algorithms and to exploit the techniques needed in building
functional quantum computers.

Currently, quantum algorithms are expressed mainly in the very low
level of quantum circuits. In the history of classical
computation, however, it was realized long time ago that
programming languages provide a technique which allows us to think
about a problem intended to solve in a high-level, conceptual way,
rather than the details of implementation. Recently, in order to
offer a similar technique in quantum computation, some authors
begun to study the design and semantics of quantum programming
languages. In the pool of imperative languages, the earliest
proposal for quantum programming language was made by Knill
in~\cite{Kn96}, where a set of basic principles for writing
quantum pseudo-code was outlined and an imperative pseudo-code
suitable for implementation on a quantum random access machine was
defined. The first real quantum programming language, QCL, was
proposed and a simulator for this language was implemented by
\"{O}mer~\cite{Om98, Om03}. A quantum programming language in the
style of Dijkstra's guarded-command language, qGCL, is designed by
Sanders and Zuliani in~\cite{Sa00, Zu01, Zu05a, Zu05b}. A
probabilistic predicate transformer semantics of qGCL was given, a
refinement calculus for it was introduced, and a compiler from
qGCL to a simple quantum architecture was defined. A quantum
extension of C++ was also proposed by Bettelli et al~\cite{Be03},
and it was implemented in the form of a C++ library. In the
functional programming style, the first quantum language, QFC, was
defined by Selinger~\cite{Se04} based on the idea of classical
control and quantum data. Programs in the language QFC are
represented via a functional version of flow charts, and QFC has a
denotational semantics in terms of complete partial orders of
super-operators. In addition, quantum process calculus CQP
(Communicating Quantum Processes) was introduced by Gay and
Nagarajan~\cite{GN05, GN06}, and QPAlg (Quantum Process Algebra)
was proposed by Jorrand and Lalire~\cite{JL04, LJ04} in order to
support the formal specification and verification of quantum
cryptographic protocols. Also, Feng et al \cite{Fe06} defined a
model qCCS of quantum processes, which is a natural quantum
extension of classical value-passing CCS with the input and output
of quantum states, and unitary transformations and measurements on
quantum systems. Semantic techniques for quantum computation have
also been investigated in some abstract, language-independent
ways. For example, a notion of quantum weakest precondition is
introduced and a Stone-type duality between the state transition
semantics and the predicate transformer semantics for quantum
programs is established by D'Hondt and Panangaden \cite{DP06}, and
proof rules for probabilistic programs were generalized by Feng et
al ~\cite{Fe05} to purely quantum programs. There are already two
excellent surveys on quantum programming languages and related
researches~\cite{Se04a, Ga06}

Loop is a powerful program construct in classical computation
\cite{Di76}. In the area of quantum computation, looping technique
has also attracted a few authors' attention. For example, Bernstein
and Vazirani~\cite{BV93, BV97} introduced some programming
primitives including looping in the context of quantum Turing
machines; some high-level control features such as loop and
recursion are provided in Selinger's functional quantum programming
language QFC. However, the full power of quantum loop programs is
still to be exploited. The exploitation of such power definitely
requires a deep understanding of the mechanism of quantum loops. The
purpose of this paper is to examine thoroughly mechanism of quantum
loops in a language-independent way, and to give some convenient
criteria for deciding termination of a general quantum loop on a
given input.

This paper is organized as follows. Section 2 is a preliminary
section in which some basic notions from quantum mechanics needed in
this paper are reviewed. In Section 3, a general scheme of quantum
loop programs is introduced, the computational process of a quantum
loop is described, and the essential difference between quantum
loops and classical loops is analyzed. In addition, we introduce the
notions of termination and almost termination of a quantum loop. The
function computed by a quantum loop is also defined. Quantum walks
are considered to show the expressive power of quantum loops. In
Section 4, we find a necessary and sufficient condition under which
a quantum loop program terminates on a given mixed input state
(Theorem~\ref{prop: 5.1}). In Section 5, a similar condition is
given for almost termination (Theorem~\ref{thm: 2}). Furthermore, we
prove that a quantum loop is almost terminating if and only if it is
uniformly almost terminating (Theorem~\ref{thm:teralter}), and a
small disturbance either on the unitary transformation in the loop
body (Theorem~\ref{thm:disturbU}) or on the measurement in the loop
guard (Theorem~\ref{thm:disturbM}) can make any quantum loop
(almost) terminating. In Section 6, a representation of the function
computed by a quantum loop is presented in terms of finite
summations of complex matrices (Theorem~\ref{prop: 6.4}). To
illustrate the notions and results presented in the previous
sections, Sections 7 is devoted to some examples which observe the
computational behavior of two simplest classes of quantum loops: one
qubit loops, and two qubit loops defined by controlled operations.
Section 8 is the concluding section in which we draw the conclusion
and point out some problems for further studies.

\section{Preliminaries}

For convenience of the reader we briefly recall some basic notions
from quantum theory and fix the notations needed in the sequel. We
refer to \cite{NC00} for more details.

An isolated physical system is associated with a Hilbert space which
is called the state space of the system. We only need to consider
finite dimensional Hilbert space in quantum computation. An
$n-$dimensional Hilbert space is an $n-$dimensional complex vector
space $H$ together with an inner product which is a mapping
$\langle\cdot|\cdot\rangle:H\times H\rightarrow \mathbf{C}$
satisfying the following properties: \begin{enumerate} \item
$\langle\phi|\phi\rangle\geq 0$ with equality if and only if
$|\phi\rangle =0$; \item
$\langle\phi|\psi\rangle=\langle\psi|\phi\rangle^{\ast}$; \item
$\langle\phi|\lambda_1\psi_1+\lambda_2\psi_2\rangle=
\lambda_1\langle\phi|\psi_1\rangle+\lambda_2\langle\phi|\psi_2\rangle$,\end{enumerate}
where $\mathbf{C}$ is the set of complex numbers, and
$\lambda^{\ast}$ stands for the conjugate of $\lambda$ for each
complex number $\lambda\in \mathbf{C}$. For any vector
$|\psi\rangle$ in $H$, its length $|||\psi\rangle||$ is defined to
be $\sqrt{\langle\psi|\psi\rangle}$. Let $V$ be a set of vectors in
a Hilbert space $H$. Then $span(V)$ is defined to be the subspace of
$H$ spanned by $V$, that is, it consists of all linear combinations
of vectors in $V$. An orthonormal basis of a Hilbert space $H$ is a
basis $\{|i\rangle\}$ with $\langle i|j\rangle=\begin{cases}1, &
\mbox{if }i=j,\\ 0, & \mbox{otherwise}.\end{cases}$ Then the trace
of a linear operator $A$ on $H$ is defined to be
$tr(A)=\sum_{i}\langle i|A|i\rangle.$

A pure state of a quantum system is a unit vector in its state
space, that is, a vector $|\psi\rangle$ with $|||\psi\rangle||=1$,
and a mixed state is represented by a density operator. A density
operator in a Hilbert space $H$ is a linear operator $\rho$ on it
fulfilling the following conditions:
\begin{enumerate}\item $\rho$ is positive in the sense that $\langle
\psi|\rho|\psi\rangle \geq 0$ for all $|\psi\rangle$; \item
$tr(\rho)=1$.\end{enumerate} An equivalent concept of density
operator is ensemble of pure states. An ensemble is a set of the
form $\{(p_i,|\psi_i\rangle)\}$ such that $p_i \geq 0$ and
$|\psi_i\rangle$ is a pure state for each $i$, and $\sum_{i}p_i=1$.
Then $\rho=\sum_{i}p_i|\psi_i\rangle\langle\psi_i|$ is a density
operator, and conversely each density operator can be generated by
an ensemble of pure states in this way. The set of density operators
on $H$ is denoted $\mathcal{D}(H)$. A positive operator $A$ is
called a partial density operator if $tr(A)\leq 1$. We write
$\mathcal{D}^{-}(H)$ for the set of partial density operators on
$H$. Obviously, $\mathcal{D}(H)\subseteq \mathcal{D}^{-}(H)$.

The evolution of a closed quantum system is described by a unitary
operator on its state space. A linear operator $U$ on a Hilbert
space $H$ is said to be unitary if $U^{\dag}U=I_H$, where $I_H$ is
the identity operator on $H$, and $U^{\dag}$ is the adjoint of $U$.
If the states of the system at times $t_1$ and $t_2$ are $\rho_1$
and $\rho_2$, respectively, then $\rho_2=U\rho_1U^{\dag}$ for some
unitary operator $U$ which depends only on $t_1$ and $t_2$. In
particular, if $\rho_1$ and $\rho_2$ are pure states
$|\psi_1\rangle$ and $|\psi_2\rangle$, respectively, that is,
$\rho_1=|\psi_1\rangle\langle \psi_1|$ and
$\rho_2=|\psi_2\rangle\langle \psi_2|$, then we have $|\psi_2\rangle
=U|\psi_1\rangle$.

A quantum measurement is described by a collection $\{M_m\}$ of
measurement operators, where the indexes $m$ refer to the
measurement outcomes. It is required that the measurement operators
satisfy the completeness equation $\sum_{m}M_m^{\dag}M_m=I_H$. If
the system is in state $\rho$, then the probability that measurement
result $m$ occurs is given by $p(m)=tr(M_m^{\dag}M_m\rho)$, and the
state of the system after the measurement is $\frac{M_m\rho
M_m^{\dag}}{p(m)}.$ For the case that $\rho$ is a pure state
$|\psi\rangle$, that is, $\rho=|\psi\rangle\langle \psi|$, we have
$p(m)=||M_m|\psi\rangle||^{2}$, and the post-measurement state is
$\frac{P_m|\psi\rangle}{\sqrt{p(m)}}$. In particular, a projective
measurement is described by an observable which is represented by a
Hermitian operator. A Hermitian operator is a linear operator $M$
with $M^{\dag}=M$. An eigenvector of a linear operator $A$ is a
nonzero vector $|\lambda\rangle$ such that $A|\lambda\rangle
=\lambda |\lambda\rangle$ for some $\lambda\in \mathbf{C}$, where
$\lambda$ is called the eigenvalue of $A$ corresponding to
$|\lambda\rangle$. We write $spec(A)$ for the set of eigenvalues of
$A$ which is called the spectrum of $A$. It is well known that all
eigenvalues of a Hermitian operator $M$ are reals. Let $M=\sum_{m\in
spec(M)}mP_m$ be the spectral decomposition of $M$ where for each
$m\in spec(M)$, $P_m$ is the projector to its corresponding
eigenspace. Obviously, these projectors form a quantum measurement
$\{P_m:m\in spec(M)\}$. If the state of a quantum system is $\rho$,
then the probability that result $m$ occurs when measuring $M$ on
the system is $p(m)=tr(P_m\rho),$ and the post-measurement state of
the system is $\frac{P_m\rho P_m}{p(m)}.$

The state space of a composite system is the tensor product of the
state spaces of its components. Let $H_1$ and $H_2$ be two Hilbert
spaces. Then their tensor product $H_1\otimes H_2$ consists of
linear combinations of vectors $|\psi_1\psi_2\rangle
=|\psi_1\rangle\otimes |\psi_2\rangle$ with $|\psi_1\rangle\in
H_1$ and $|\psi_2\rangle\in H_2$. For any linear operator $A_1$ on
$H_1$ and $A_2$ on $H_2$, $A_1\otimes A_2$ is an operator on
$H_1\otimes H_2$ and it is defined by $$(A_1\otimes
A_2)|\psi_1\psi_2\rangle = A_1|\psi_1\rangle\otimes
A_2|\psi_2\rangle$$ for each $|\psi_1\rangle \in H_1$ and
$|\psi_2\rangle \in H_2$. Since density operators are special
linear operators, their tensor product is then well-defined. If
component system $i$ is in state $\rho_i$ for each $i$, then the
state of the composite system is $\bigotimes_i\rho_i$.

\section{Basic Definitions}

We first give a general and formal formulation of quantum loop
programs. Suppose that we have $n$ quantum registers
$q_1,\dots,q_n$, and their state spaces are $H_1,\dots,H_n$,
respectively. We further assume that $U$ is a unitary operator on
the tensor product space $H=\bigotimes_{i=1}^{n}H_i.$ Let
$M=\sum_{m}mP_m$ be a projective measurement on $H$. Then for any
$X\subseteq spec(M)$, the quantum loop program defined by $U$, $M$
and $X$ may be written as follows:
\begin{equation}\label{eq: lp1}\mathbf{while}\ (M\in X)\
\{\overline{q}:=U\overline{q}\}\end{equation} where $\overline{q}$
is used to denote the sequence $q_1,\dots,q_n$ of quantum registers.
Let $P_X=\sum_{m\in X}P_m$ and $P_{\overline{X}}=I_H-P_X=\sum_{m\in
spec(M)-X}P_m,$ where $I_H$ is the unit operator on $H$. Then the
guard \textquotedblleft$M\in X$\textquotedblright of loop~(\ref{eq:
lp1}) means that the projective measurement
$\{P_X,P_{\overline{X}}\}$ is applied to $\overline{q}$, and the
outcome corresponding to $P_X$ is observed. The body of the loop is
the assignment
\textquotedblleft$\overline{q}:=U\overline{q}$\textquotedblright,
that is, a command of performing unitary transformation $U$ on the
state of the sequence $\overline{q}$ of quantum registers. This loop
can be visualized by Fig.1.

It is worth noting that the projective measurement we perform to
check the guard condition of loop~(\ref{eq: lp1}) is
$\{P_X,P_{\overline{X}}\}$ rather than $M$ itself, because we need
only tell whether or not the measurement outcome belongs to $X$. Any
further information about the exact outcome is useless, and will
bring unnecessary disturbance to the system we measured.

\begin{figure}\centering
\includegraphics{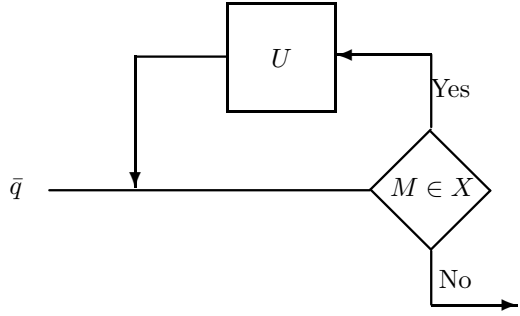}
\caption{Quantum loop~(\ref{eq: lp1})} \label{fig 1 }
\end{figure}

We now examine the computational process of the above loop program.
For any input state $\rho_0=\rho\in \mathcal{D}(H)$, the behavior of
the above quantum loop can be described in the following unwound way
(see Fig.2):

\begin{figure}\centering
\includegraphics{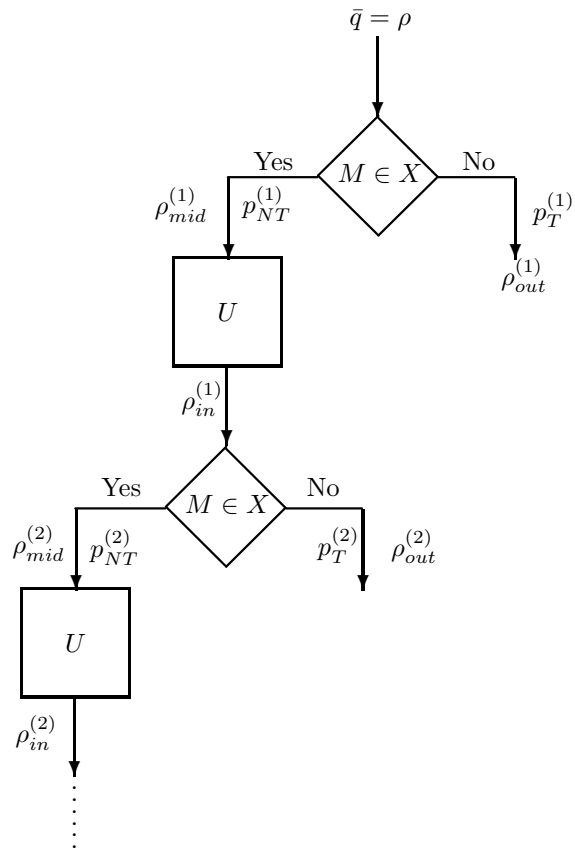}
\caption{The computational process of loop~(\ref{eq:
lp1})}\label{fig 2 }
\end{figure}

\begin{enumerate}
\item This is the initial step. The loop program performs the
projective measurement $\{P_X,P_{\overline{X}}\}$ on the input state
$\rho$. If the outcome corresponding to $P_X$ is observed, then the
program performs the given unitary operation $U$ on the
post-measurement state. Otherwise the program terminates. Formally,
the loop will terminate with probability
$p_T^{(1)}(\rho)=tr(P_{\overline{X}}\rho)$ and it will continue with
probability $p_{NT}^{(1)}(\rho)=1-p_T^{(1)}(\rho)=tr(P_X\rho).$ In
the case of termination, the output of the loop is
$$\rho_{out}^{(1)}=\frac{P_{\overline{X}}\rho
P_{\overline{X}}}{p_T^{(1)}(\rho)},$$ and in the case of
nontermination, the state of $\overline{q}$ system after the
measurement is $$\rho_{mid}^{(1)}=\frac{P_X\rho
P_X}{p_{NT}^{(1)}(\rho)}.$$ Furthermore, $\rho_{mid}^{(1)}$ will be
fed to the unitary operation $U$ and then the state
$\rho_{in}^{(1)}=U\rho_{mid}^{(1)}U^{\dag}$ is returned, which will
be used as the input state in the next step. \item This is the
induction step. Suppose that the loop has performed $n$ steps, and
it did not terminate at the $n$th step, that is, $p_{NT}^{(n)}>0$.
If $\rho_{in}^{(n)}$ is the state of $\overline{q}$ system returned
at the $n$th step, then in the $(n+1)$th step, the termination
probability is
$p_T^{(n+1)}(\rho)=tr(P_{\overline{X}}\rho_{in}^{(n)})$ and the
output is
$$\rho_{out}^{(n+1)}=\frac{P_{\overline{X}}\rho_{in}^{(n)}P_{\overline{X}}}{p_T^{(n+1)}(\rho)}.$$
The loop continues to perform the unitary operation $U$ on the
post-measurement state
$$\rho_{mid}^{(n+1)}=\frac{P_{X}\rho_{in}^{(n)}P_X}{p_{NT}^{(n+1)}(\rho)}$$
with probability
$p_{NT}^{(n+1)}(\rho)=1-p_T^{(n+1)}(\rho)=tr(P_X\rho_{in}^{(n)}),$
and the state $\rho_{in}^{(n+1)}=U\rho_{mid}^{(n+1)}U^{\dag}$ will
be returned.\end{enumerate}

Note that not only a pure quantum state but also a mixed state is
allowed to feed into a quantum loop. In fact, quantum programming
with mixed states has already been considered in the previous
literature; for example, see~\cite{Se04, Zu05b}.

There is an essential difference between the computing process of
quantum loops and that of classical loops. In a classical loop the
states of variables do not change during verification of the loop
condition. However, in a quantum loop it is impossible to check the
loop condition directly. Instead, the loop program needs to extract
information about the registers $q_1,\dots,q_n$ by performing a
measurement $M$ and thus their states will be changed.

To demonstrate the expressive power of quantum loops, let us
consider an interesting example. Quantum walk is a natural quantum
extension of classical random walk, which in turn has proved to be a
fundamental tool in computer science, especially in the designing of
algorithms \cite{KS83}. In this example, we consider a discrete
coined quantum walk on an $n$-cycle with an absorbing boundary at
position 1, and express this kind of quantum walk in our language of
quantum loops. For more details about quantum walk on a cycle, or
more generally, on any graph, we refer to \cite{AAK+01}. The
following example shows that a quantum walk can be described very
well in the language of quantum loops.

\begin{exam}\label{exam} Let $H_A$ be a 2-dimensional `coin' space with orthonomal basis
states $|0\>$ and $|1\>$, and $H_V$ be the $n$-dimensional principle
space spanned by the position vectors $|i\>:i=0,\dots,n-1$. Then
each step of the quantum walk we are concerned with consists of
three sub-steps:
\begin{enumerate}
\item A `coin-tossing operator' $H=|+\>\<0|+|-\>\<1|$ is
applied to the coin space, where $|+\>=(|0\>+|1\>)/\sqrt{2}$ and
$|-\>=(|0\>-|1\>)/\sqrt{2}$.

\item A shift operator
$$S=\sum_{i=0}^{n-1}|i\ominus 1\>\<i|\otimes |0\>\<0|+
\sum_{i=0}^{n-1}|i\oplus 1\>\<i|\otimes |1\>\<1|$$ is performed on
the space $H_V\otimes H_A$, which makes the quantum walk one step
left or right according to the coin state. Here $\ominus$ and
$\oplus$ denote subtraction and addition modulo $n$, respectively.

\item Measure the principle system to see if the current
position of the walk is $1$. If the answer is `yes' then terminate
the walk, otherwise the walk continues.
\end{enumerate}
Formally, we can formulate the walk described above by a quantum
loop:
\begin{equation}\label{lp:qwalk}
    \mathbf{while}\ (M\neq 1)\
\{\overline{q}:=U\overline{q}\}
\end{equation}
where $M=\sum_{i=0}^{n-1} i|i\>\<i|\otimes I_2$, $U=S(I_n\otimes
H)$, and $\overline{q}$ is a quantum register in $H_V\otimes H_A$.
For simplicity, we write $M\neq 1$ in the loop guard to denote $M\in
X$ for $X=\{0,2,\dots,n-1\}$.
\end{exam}

One of the most important problems concerning the behavior of a loop
program is its termination.

\begin{defn}\label{defn: 1}
\begin{enumerate}
\item
If $p_{NT}^{(n)}(\rho)=0$ for some positive integer $n$, then it is
said that the loop with input $\rho$ terminates.
\item
The nonterminating probability  of the loop with input $\rho$ is
defined to be
$$p_{NT}(\rho)=\lim_{n\rightarrow\infty}p_{NT}^{n+}(\rho)$$
where (and in the sequel)
$p_{NT}^{n+}(\rho)=\prod_{i=1}^{n}p_{NT}^{(i)}(\rho)$ denotes the
probability that the loop does not terminate after $n$ steps.

\item We
say that the loop with input $\rho$ almost terminate whenever
$p_{NT}(\rho)=0$.
\item
If $p_{NT}(\rho)>0$, then we say that the loop with input $\rho$
does not terminate. \end{enumerate} \end{defn}

Intuitively, a quantum loop almost terminates if for any
$\epsilon>0$, there exists a big enough positive integer
$n(\epsilon)$ such that the probability that the loop terminates at
the $n(\epsilon)$th step is greater than $1-\epsilon$. Obviously, if
a quantum loop terminates on a given input state, then it also
almost terminates on the same input.

A classical loop may terminate or not, but a quantum loop has an
additional possibility of almost termination. Clearly, this is
caused by the probabilistic nature of quantum mechanics.

\begin{defn}\label{defn: 2}\begin{enumerate}\item A quantum loop program is said
to be terminating (resp. almost terminating) if it terminates (resp.
almost terminates) with all input $\rho\in \mathcal{D}(H)$.
\item A quantum loop is uniformly almost terminating if for any $\epsilon>0$ there exists a
positive integer $n(\epsilon)$ such that
$p_{NT}^{n+}(\rho)<\epsilon$ holds for all input $\rho\in
\mathcal{D}(H)$ whenever $n\geq n(\epsilon)$.
\end{enumerate}\end{defn}

It is clear that uniformly almost terminating quantum loops are
almost terminating.

Note that the case of $X=\emptyset$ or $spec(M)$ is trivial. In
fact, the loop~(\ref{eq: lp1}) is equivalent to $$\mathbf{while}\
(false)\ \{\overline{q}:=U\overline{q}\}$$ when $X=\emptyset$, and
it is equivalent to $$\mathbf{while}\ (true)\
\{\overline{q}:=U\overline{q}\}$$ when $X=spec(M)$. The former
terminates immediately and does nothing, and the latter will loop
forever. In what follows we always assume that $\emptyset \subset
X\subset spec(M)$.

In the computational process of a loop program, a density operator
is input, and a density operator is outputted with a certain
probability at each step. Thus, we have to synthesize these density
operators returned at all steps according to the respective
probabilities into a single one as the overall output. Note that
sometimes the loop does not terminate with a nonzero probability.
The synthesized output may not be a density operator but only a
partial density operator. Then a loop defines a function from
density operators to partial density operators on $H$.

\begin{defn}\label{defn: 3} The loop~(\ref{eq: lp1}) defines a function
$F:\mathcal{D}(H)\rightarrow \mathcal{D}^{-}(H)$ in the following
way:
$$F(\rho)=\rho_{out}=\sum_{n=1}^{\infty}p_{NT}^{(n-1)+}(\rho)p_{T}^{(n)}(\rho)\rho_{out}^{(n)}$$
for each $\rho\in \mathcal{D}(H).$  The function $F$ is called the
function computed by the loop~(\ref{eq: lp1}).\end{defn}

It should be noted that in the defining equation of $F(\rho)$ the
quantity $p_{NT}^{(n-1)+}(\rho)p_{T}^{(n)}(\rho)$ is the probability
that the loop does not terminate at steps from $1$ to $n-1$ but it
terminates at the $n$th step.

For the case that $\rho$ is a pure state, that is,
$\rho=|\psi\rangle\langle\psi|$ for some pure state $|\psi\rangle$,
we will write $F(|\psi\rangle)$ in place of $F(\rho)$ for
simplicity.

In the remainder of this section, we are going to present some basic
properties of quantum loops. For any operator $A$ on $H$, we write
$A_X=P_XAP_X$, that is, $A_X$ is the restriction of $A$ on the
subspace of $H$ corresponding to the projector $P_X$. First, the
computational process of quantum loop~(\ref{eq: lp1}) can be
summarized as:

\begin{lem}\label{lem: 1} Let $\rho$ be the input state to the loop~(\ref{eq: lp1}). Then for any positive integer $n$, we
have:
\begin{equation} \label{eqn:pn}
  p_{NT}^{n+}(\rho)= tr(U_X^{n-1} \rho_X U_X^{\dag n-1})
\end{equation}
and
\begin{equation}\label{eqn:frho}
  F(\rho)
  = P_{\overline{X}}\rho P_{\overline{X}} + \displaystyle P_{\overline{X}}U \left(\sum_{n=0}^{\infty} U_X^n \rho_X U_X^{^\dag n}\right) U^\dag
  P_{\overline{X}}.
\end{equation}
\end{lem}
{\it Proof.} First, it is easy to check by induction on $n$ that
\begin{equation}\label{eqn:nt}
    p_{NT}^{(n)}(\rho)=\begin{cases}\displaystyle\frac{tr(U_X^{n-1} \rho_X U_X^{\dag n-1})}{tr(U_X^{n-2} \rho_X U_X^{\dag n-2})}, & \mbox{if  } n
    \geq 2,\\ \\
tr(\rho_X), & \mbox{if  } n=1,\end{cases}
\end{equation}
\begin{equation}\label{eqn:t}
    p_{T}^{(n)}(\rho)=\begin{cases}\displaystyle\frac{tr(P_{\overline{X}}UU_X^{n-2}
    \rho_X U_X^{\dag n-2}U^{\dag})}{tr(U_X^{n-2} \rho_X U_X^{\dag n-2})}, & \mbox{if  } n
    \geq 2,\\ \\
1-tr(\rho_X), & \mbox{if  } n=1,\end{cases}
\end{equation}
and
\begin{equation}\label{eqn:out}
    \rho_{out}^{(n)}=\frac{P_{\overline{X}}UU_X^{n-2} \rho_X U_X^{\dag n-2}U^{\dag}P_{\overline{X}}}{tr(P_{\overline{X}}UU_X^{n-2}
    \rho_X U_X^{\dag n-2}U^{\dag})}.
\end{equation}
Then Eq.~(\ref{eqn:pn}) follows from Eq.~(\ref{eqn:nt}), and
Eq.~(\ref{eqn:frho}) comes from Eqs.~(\ref{eqn:pn}), (\ref{eqn:t}),
and (\ref{eqn:out}). \hfill $\square$

\vspace{1em}

 It should be pointed out here that convergence of the
infinite series in Definition~\ref{defn: 3} and Eq.~(\ref{eqn:frho})
is guaranteed by the facts that the set $\mathcal{D}^-(H)$ is a
directed complete poset under the L\"{o}wner order and the sequence
$\left\{\sum_{n=0}^{k} U_X^n \rho_X U_X^{^\dag
n}\right\}_{k=0}^\infty$ is non-decreasing in this order. For the
details, we refer to \cite{Se04}. On the other hand, from
Eq.~(\ref{eqn:frho}) and Kraus representation theorem (\cite{NC00},
Theorem 8.1) we notice that the function $F$ computed by loop
(\ref{eq: lp1}) is a super-operator (also called quantum operation).

Let $H_X$ be the subspace of $H$ with projector $P_X$, and
$H_{\overline{X}}$ the subspace with projector $P_{\overline{X}}$.
The following proposition clarifies the range of the function $F$
computed by the loop~(\ref{eq: lp1}).

\begin{prop}\label{prop: 1} For each $\rho\in \mathcal{D}(H)$, we have:
\begin{enumerate}\item $\langle \phi|F(\rho)|\psi\rangle =0$ if
$|\phi\rangle$ or $|\psi\rangle\in H_X$; \item
$tr(F(\rho))=1-p_{NT}(\rho)$. Thus, $F(\rho)\in \mathcal{D}(H)$ if
and only if the loop~(\ref{eq: lp1}) with input state $\rho$ almost
terminates.\end{enumerate}\end{prop} {\it Proof.} 1. By definition
we know that
$P_{\overline{X}}|\phi\rangle=P_{\overline{X}}|\psi\rangle=0$. Then
it follows immediately from Lemma~\ref{lem: 1}.

2. By induction on $k$ it may be shown that
$$\sum_{n=1}^{k}p_{NT}^{(n-1)+}(\rho)p_T^{(n)}(\rho)=1-p_{NT}^{k+}(\rho).$$
Then we have:
\begin{equation*}\begin{split}
tr(F(\rho)) & =
\sum_{n=1}^{\infty}p_{NT}^{(n-1)+}(\rho)p_{T}^{(n)}(\rho)\\ & =
1-\lim_{n\rightarrow
\infty}p_{NT}^{n+}(\rho)=1-p_{NT}(\rho).\end{split}\end{equation*}
The conclusion follows immediately. \hfill $\square$

\vspace{1em}

From Proposition~\ref{prop: 1}.1 we see that $F$ is indeed a
function from $\mathcal{D}(H)$ into
$\mathcal{D}^{-}(H_{\overline{X}})$.
\begin{figure*}[!t]
\begin{eqnarray}\label{eq: 8}
J_{r}(\lambda)^{N}=\left(\begin{array}{cccccc} \lambda^{N} &
\left(\begin{array}{cc}N\\
1\end{array}\right)\lambda^{N-1} & \left(\begin{array}{cc}N\\
2\end{array}\right)\lambda^{N-2} & \cdots &
\left(\begin{array}{cc}N\\
r-2\end{array}\right)\lambda^{N-r+2} &
\left(\begin{array}{cc}N\\
r-1\end{array}\right)\lambda^{N-r+1}\\
0 & \lambda^{N} & \left(\begin{array}{cc}N\\
1\end{array}\right)\lambda^{N-1} & \cdots &
\left(\begin{array}{cc}N\\
r-3\end{array}\right)\lambda^{N-r+3} &
\left(\begin{array}{cc}N\\
r-2\end{array}\right)\lambda^{N-r+2}\\
0 & 0 & \lambda^{N} & \cdots & \left(\begin{array}{cc}N\\
r-4\end{array}\right)\lambda^{N-r+4} &
\left(\begin{array}{cc}N\\
r-3\end{array}\right)\lambda^{N-r+3}\\
& & & \cdots & & \\
0 & 0 & 0 & \cdots & \lambda^{N} & \left(\begin{array}{cc}N\\
1\end{array}\right)\lambda^{N-1}\\
0 & 0 & 0 & \cdots & 0 & \lambda^{N}\end{array}\right).
\end{eqnarray}\hrulefill\vspace*{2pt}
\end{figure*}

\begin{figure*}[!t]
\begin{eqnarray}\label{eq: 9}J_r(\lambda)^{N}\mathbf{v}=(\sum_{i=0}^{r-1} \left(\begin{array}{cc}N\\
i\end{array}\right)\lambda^{N-i}v_{i+1}, \sum_{i=0}^{r-2}\left(\begin{array}{cc}N\\
i\end{array}\right)\lambda^{N-i}v_{i+2}, \cdots,\lambda^{N}v_{r-1}+\left(\begin{array}{cc}N\\
1\end{array}\right)\lambda^{N-1}v_r,\lambda^{N}v_r)^{T}.\end{eqnarray}\hrulefill\vspace*{2pt}
\end{figure*}

\section{Termination}

The aim of this section is to give a necessary and sufficient
condition under which the loop~(\ref{eq: lp1}) terminates on a given
input state.

We first give a lemma which allows us to decompose an input density
matrix into a sequence of simpler input density matrices when
examining termination of a quantum loop.

\begin{lem}\label{prop: 6.3} Let $\rho=\sum_i p_i\rho_i$
where $p_i>0$ and $\rho_i\in \mathcal{D}(H)$ for each $i$, and
$\sum_{i}p_i=1$. Then the loop~(\ref{eq: lp1}) with input $\rho$
terminates if and only if it terminates with input $\rho_i$ for all
$i$.\end{lem} {\it Proof.} For each $i$, if the loop~(\ref{eq: lp1})
with input $\rho_i$ terminates, then there exists a positive integer
$n_i$ such that $p_{NT}^{n_i+}(\rho_i)=0$. Let $n_0=\max_{i}n_i$.
Then $p_{NT}^{n_0+}(\rho_i)=0$ for all $i$, and this yields
$$p_{NT}^{n_0+}(\rho)=\sum_{i} p_ip_{NT}^{n_0+}(\rho_i) =0.$$

Conversely, if the loop~(\ref{eq: lp1}) with input $\rho$
terminates, then there exists a positive integer $n_0$ such that
$p_{NT}^{n_0+}(\rho)=0$. This implies that for each $i$,
$p_{NT}^{n_0+}(\rho_i)=0$ because $p_{NT}^{n_0+}(\rho_i)\geq 0$ for
each $i$. \hfill $\square$

\vspace{1em}

If $\{(p_i,|\psi_i\rangle)\}$ is an ensemble with $p_i>0$ for all
$i$, and $\rho=\sum_{i}p_i|\psi_i\rangle\langle\psi_i|,$ then the
above lemma asserts that the loop~(\ref{eq: lp1}) terminates on
input mixed state $\rho$ if and only if it terminates on input pure
state $|\psi_i\rangle$ for all $i$. In particular, we have:

\begin{cor}\label{cor: 6.1} A quantum loop is terminating if and only if it terminates with all pure input states. \end{cor}

Second, the termination problem of a quantum loop may be reduced to
a corresponding problem of a classical loop in the field of complex
numbers. Let $|m_1\>,|m_2\>,\dots,|m_K\>$ be an orthonormal basis of
$H$ such that
$$\sum_{i=1}^k |m_i\>\<m_i| = P_X, \sum_{i=k+1}^K |m_i\>\<m_i| =
P_{\overline{X}},$$ where $1\leq k\leq K$. Without any loss of
generality, we assume in the sequel that the matrix representations
of $U, U_X, \rho_X$ (denoted also by $U, U_X, \rho_X$ respectively
for simplicity) are taken according to this basis. In particular,
for each pure state $|\psi\>$ we write $|\psi\>_X$ for the vector
representation of $P_X|\psi\>$ under this basis.

\begin{lem}\label{prop: 6.2} The quantum loop~(\ref{eq: lp1}) terminates with input
$\rho\in \mathcal{D}(H)$ if and only if $U_X^{N}\rho_XU^{\dag
N}_X=\mathbf{0}_{k\times k}$ for some nonnegative integer $N$, where
$\mathbf{0}_{k\times k}$ is the $(k\times k)$-zero matrix. In
particular, it terminates with pure input state $|\psi\rangle$ if
and only if $U_X^{N}|\psi\rangle_X=\mathbf{0}$ for some nonnegative
integer $N$, where $\mathbf{0}$ is the zero vector of length
$k$.\end{lem} {\it Proof.} This result follows from
Eq.~(\ref{eqn:pn}) and the fact that $tr(A)=\mathbf{0}$ if and only
if $A=\mathbf{0}$ when $A$ is positive semi-definite. \hfill
$\square$

\vspace{1em}

Third, we show certain invariance of termination of a classical loop
under a nonsingular transformation.

\begin{lem}\label{lem: 6.1} Let $S$ be a nonsingular $(k\times
k)-$complex matrix. Then the (classical) loop:
\begin{equation*}\label{eq: lp7}\mathbf{while}\ (\mathbf{v}\neq
\mathbf{0})\ \{\mathbf{v}:=U_X\mathbf{v}\}\ \ (\mathbf{v}\in
\mathbf{C}^{k})\end{equation*} terminates on input
$\mathbf{v}_0\in \mathbf{C}^{k}$ if and only if the following
loop:
$$\mathbf{while}\ (\mathbf{v}\neq \mathbf{0})\
\{\mathbf{v}:=(SU_XS^{-1})\mathbf{v}\}\ \ (\mathbf{v}\in
\mathbf{C}^{k})$$ terminates on input $S\mathbf{v}_0$.\end{lem} {\it
Proof.} Note that $S\mathbf{v}\neq \mathbf{0}$ if and only if
$\mathbf{v}\neq \mathbf{0}$ because $S$ is nonsingular. Then the
conclusion follows from a simple calculation. \hfill $\square$

\vspace{1em}

Furthermore, we shall need the famous Jordan normal form theorem in
the proof of the main result in this section.

\begin{lem}\label{lem: 6.2} (Jordan normal form; \cite{Bh91}) For
any $(k\times k)-$complex matrix $A$, there is a nonsingular
$(k\times k)-$complex matrix $S$ such that $A=SJ(A)S^{-1}$, where
\begin{equation*}\begin{split}J(A) &
=diag(J_{k_1}(\lambda_1),J_{k_2}(\lambda_2),\dots,J_{k_l}(\lambda_l))\\
& =\left(\begin{array}{ccccc}J_{k_1}(\lambda_1) &  &  &  & \\
 & J_{k_2}(\lambda_2) &  &  & \\
& & \ddots &  & \\
 &  &  & \ddots & \\
 &  &  &  & J_{k_l}(\lambda_l)
\end{array}\right)\end{split}\end{equation*} is
the Jordan normal form of $A$, $\sum_{i=1}^{l}k_i=k$, and
\begin{equation}J_{k_i}(\lambda_i)=\left(\begin{array}{ccccc}\lambda_i & 1 &  &  & \\
 & \lambda_i & 1 &  & \\
& & \ddots & \ddots & \\
 &  &  & \ddots & 1\\
 &  &  &  & \lambda_i
\end{array}\right).\end{equation}is a $(k_i\times k_i)$-Jordan
block for each $1\leq i\leq l$. Furthermore, if the Jordan blocks
corresponding to each distinct eigenvalue are presented in
decreasing order of the block size, then the Jordan normal form is
uniquely determined once the ordering of the eigenvalues is given.
\end{lem}

The following technical lemma is also needed.

\begin{lem}\label{lem: 6.3} Let $J_r(\lambda)$ be a $(r\times
r)-$Jordan block, and $\mathbf{v}\in \mathbf{C}^{r}$. Then
$J_r(\lambda)^{N}\mathbf{v}=\mathbf{0}$ for some nonnegative integer
$N$ if and only if $\lambda=0$ or $\mathbf{v}=\mathbf{0}$, where
$\mathbf{0}$ is the zero vector of length $r$.
\end{lem}
{\it Proof.} The \textquotedblleft if\textquotedblright part is
clear. We now prove the \textquotedblleft only if\textquotedblright
part. By a routine calculation we obtain the matrix
$J_r(\lambda)^{N}$ as in Eq.~(\ref{eq: 8}). Notice that
$J_r(\lambda)^{N}$ is an upper triangular matrix with the diagonal
entries being $\lambda^N$. So if $\lambda\neq 0$ then
$J_r(\lambda)^{N}$ is nonsingular, and then $J_r(\lambda)^{N}
\mathbf{v}=\mathbf{0}$ implies $\mathbf{v}=\mathbf{0}$. \hfill
$\square$

\vspace{1em}

Now we are able to present the main result of this section.

\begin{thm}\label{prop: 5.1} Suppose the Jordan decomposition
of $U_X$ is $U_X=SJ(U_X)S^{-1}$, where
$$J(U_X)=diag(J_{k_1}(\lambda_1),J_{k_2}(\lambda_2),\dots,J_{k_l}(\lambda_l)).$$ Let
$S^{-1}|\psi\rangle_X$ be divided into $l$ sub-vectors
$\mathbf{v}_1,\mathbf{v}_2,\dots,\mathbf{v}_l$ such that the length
of $\mathbf{v}_i$ is $k_i$. Then the loop~(\ref{eq: lp1}) terminates
on input $|\psi\rangle$ if and only if for each $1\leq i\leq l$,
$\lambda_i=0$ or $\mathbf{v}_i=\mathbf{0}$, where $\mathbf{0}$ is
the zero vector of length $k_i$.\end{thm}
{\it Proof.} Using
Lemmas~\ref{prop: 6.2} and~\ref{lem: 6.1} we know that the
loop~(\ref{eq: lp1}) terminates on input $|\psi\rangle$ if and only
if $J(U_X)^{N}S^{-1}|\psi\rangle_X=\mathbf{0}$ for some nonnegative
integer $N$. A simple calculation yields
$$\begin{array}{l}
J(U_X)^{N}S^{-1}|\psi\rangle_X\\ \\
=((J_{k_1}(\lambda_1)^{N}\mathbf{v}_1)^T,(J_{k_2}(\lambda_2)^{N}\mathbf{v}_2)^T,
\dots,(J_{k_m}(\lambda_m)^{N}\mathbf{v}_m)^T)^T.\end{array}$$
Therefore, $J(U_X)^{N}S^{-1}|\psi\rangle_X=\mathbf{0}$ for some
nonnegative integer $N$ if and only if for each $1\leq i\leq l$,
there exists a nonnegative integer $N_i$ such that
$J_{k_i}(\lambda_i)^{N_i}\mathbf{v}_i=\mathbf{0}.$ Then we complete
the proof by using Lemma~\ref{lem: 6.3}. \hfill $\square$

\begin{cor}
Loop (\ref{eq: lp1}) is terminating if and only if $U_X$ has only
zero eigenvalues.
\end{cor}

Theorem ~\ref{prop: 5.1} gives a necessary and sufficient condition
for termination of loop (\ref{eq: lp1}) on an input pure state.
Obviously, we can decide whether the loop~(\ref{eq: lp1}) terminates
on any given mixed state as input by combining Lemma~\ref{prop: 6.3}
and Theorem~\ref{prop: 5.1}. The condition for termination of loop
(\ref{eq: lp1}) can be considerably simplified in the special case
when $U_X$ is normal, that is, $U_XU_X^{\dag}=U_X^{\dag}U_X.$ In
this case, $U_X$ has the following simple spectrum decomposition:
\begin{equation}\label{eq:sdecom}
U_X=\sum_{i=1}^k \lambda_i|i\>\<i|.
\end{equation} Then from Eq.~(\ref{eqn:pn}) we have for any
$\rho\in\mathcal{D}(H)$:
\begin{eqnarray}
    p_{NT}^{n+}(\rho)&=& tr (\sum_{i,j=1}^k
    \lambda_i^{n-1}|i\>\<i|\rho|j\>\<j|\lambda_j^{*n-1})\\
    &=& \sum_{i=1}^k
    |\lambda_i|^{2(n-1)}\<i|\rho|i\>.\label{eqn:normalpnt}
\end{eqnarray} This implies immediately the following:
\begin{cor} Suppose $U_X$ is normal and its spectrum decomposition is given by Eq.~(\ref{eq:sdecom}).
Then we have:
\begin{enumerate}
  \item loop~(\ref{eq: lp1}) terminates on input state $\rho$ if and only if for any $i=1,\dots,k$,
$\lambda_i\neq 0$ implies $\<i|\rho|i\>=0$, or equivalently, $tr
(U_X\rho) =0.$
  \item loop~(\ref{eq: lp1}) is terminating if and only if $U_X =0$.
\end{enumerate}
\end{cor}

\section{Almost termination}

In this section we are going to present a necessary and sufficient
condition under which the loop~(\ref{eq: lp1}) almost terminates on
any given input state. We first give a lemma similar to
Lemma~\ref{prop: 6.3} so that a mixed input state can be reduced to
a family of pure input states.
\begin{lem}\label{prop: alpure} Let $\rho=\sum_{i}p_i\rho_i$
where $p_i>0$ and $\rho_i\in \mathcal{D}(H)$ for each $i$, and
$\sum_{i}p_i=1$. Then the loop~(\ref{eq: lp1}) with input $\rho$
almost terminates if and only if it almost terminates with input
$\rho_i$ for all $i$.\end{lem} {\it Proof.} Notice that
$p_{NT}^{n+}(\rho)=\sum_{i}p_i p_{NT}^{n+}(\rho_i)$ from
Eq.~(\ref{eqn:pn}). The result follows by taking limits about $n$ in
both sides of the above equation. \hfill $\square$

\begin{cor}\label{cor: 6.1} A quantum loop is almost terminating if and only if it almost terminates on all pure input states. \end{cor}

The following lemma is a key step in the proof of our main result in
this section.
\begin{lem}\label{lem: alpure} The loop~(\ref{eq: lp1}) almost terminates on the pure input state $|\psi\rangle$ if and only if
$\lim_{n\rightarrow\infty}||U_X^{n}|\psi\rangle||=0.$ \end{lem} {\it
Proof.} From Eq.~(\ref{eqn:pn}) we have
$p_{NT}^{n+}(|\psi\>)=\|U_X^{n-1}|\psi\>\|^2$. So
$p_{NT}(|\psi\>)=0$ if and only
$\lim_{n\rightarrow\infty}||U_X^{n}|\psi\rangle||=0$. \hfill
$\square$

\vspace{1em}

The following theorem gives a necessary and sufficient condition for
almost termination of a quantum loop on a pure input state.

\begin{thm}\label{thm: 2} Suppose that $U_X$, $S$, $J(U_X)$, $J_{k_i}(\lambda_i)$ and
$\mathbf{v}_i$ $(1\leq i\leq l)$ are given as in Theorem~\ref{prop:
5.1}. Then the loop~(\ref{eq: lp1}) almost terminates on input
$|\psi\rangle$ if and only if for each $1\leq i\leq l$,
$|\lambda_i|<1$ or $\mathbf{v}_i=\mathbf{0}$, where $\mathbf{0}$ is
the zero vector of length $k_i$.\end{thm} {\it Proof.} First, for
any nonnegative integer $n$, we have $U_X^{n}|\psi\rangle =
SJ(U_X)^{n}S^{-1}|\psi\rangle.$ Then
$\lim_{n\rightarrow\infty}||U_X^{n}|\psi\rangle||=0$ if and only if
\begin{equation}\label{eq: 11}\lim_{n\rightarrow\infty}||J(U_X)^{n}S^{-1}|\psi\rangle||=0\end{equation}
since $S$ is nonsingular. Using Lemma~\ref{lem: alpure} we know that
the loop~(\ref{eq: lp1}) almost terminates if and only if
Eq.~(\ref{eq: 11}) holds. Note that
$$
\begin{array}{l}
J(U_X)^{n}S^{-1}|\psi\rangle\\
=((J_{k_1}(\lambda_1)^{n}\mathbf{v}_1)^T,
(J_{k_2}(\lambda_2)^{n}\mathbf{v}_2)^T,\dots,(J_{k_l}(\lambda_l)^{n}\mathbf{v}_l)^T)^T.\end{array}$$
Then Eq.~(\ref{eq: 11}) holds if and only if
\begin{equation}\label{eq:
12}\lim_{n\rightarrow\infty}||J_{k_i}(\lambda_i)\mathbf{v}_i||=0\end{equation}
for all $1\leq i\leq l$. Furthermore, for each $1\leq i\leq l$, from
Eq.~(\ref{eq: 9}) we see that Eq.~(\ref{eq: 12}) holds if and only
if the following $k_i$ equations are valid:
\begin{equation}\label{eq:case}\begin{cases}
\lim_{n\rightarrow\infty}\sum_{j=0}^{k_i-1} \left(\begin{array}{cc}n
\\ j\end{array}\right)\lambda_i^{n-j}v_{i(j+1)}=0,\\
 \lim_{n\rightarrow\infty}\sum_{j=0}^{k_i-2}\left(\begin{array}{cc}n\\
j\end{array}\right)\lambda_i^{n-j}v_{i(j+2)}=0,\\
\dots\dots\dots\dots\dots\dots\\
\lim_{n\rightarrow\infty}[\lambda_i^{n}v_{i(k-1)}+\left(\begin{array}{cc}n\\
1\end{array}\right)\lambda_i^{n-1}v_{ik}]=0,\\
\lim_{n\rightarrow\infty}\lambda_i^{n}v_{ik}=0,\end{cases}\end{equation}
where it is assumed that
$\mathbf{v}_i=(v_{i1},v_{i2},\dots,v_{ik_i})$.

If $|\lambda_i|<1$, then $\lim_{n\rightarrow\infty}\left(\begin{array}{cc}n\\
j\end{array}\right)\lambda_{i}^{n-j}=0$ for any $0\leq j\leq k_i-1$,
and all of the above equations in Eq.~(\ref{eq:case}) follow. On the
other hand, if $|\lambda_i|\geq 1,$ then from the last equation in
Eq.~(\ref{eq:case}) we know that $v_{ik}=0$. Putting $v_{ik}=0$ into
the second equation from bottom in Eq.~(\ref{eq:case}) we obtain
$v_{i(k-1)}=0$. We can further move from bottom to top in
Eq~(\ref{eq:case}) in this way, and finally we get
$v_{i1}=v_{i2}=\dots=v_{i(k-1)}=v_{ik}=0$. This completes the proof.
\hfill $\square$

\begin{cor}\label{cor:normlessthan1}
Loop (\ref{eq: lp1}) is almost terminating if and only if all the
eigenvalues of $U_X$ have norms less than 1.
\end{cor}

In the case when $U_X$ is normal, we have the following corollary
which is also easy to prove directly from Eq.~(\ref{eqn:normalpnt}).
\begin{cor} Suppose $U_X$ is normal and Eq.~(\ref{eq:sdecom}) is its spectrum
decomposition. Then
\begin{enumerate}
  \item loop~(\ref{eq: lp1}) with input $\rho$ almost terminates if and only if  for any $i=1,\dots,k$,
$|\lambda_i|=1$ implies $\<i|\rho|i\>=0$, i.e., the set
\begin{equation*}
    I' \equiv \{\ 1\leq i\leq k\ |\ |\lambda_i|=1\ \mbox{  and  }\ \<i|\rho|i\>
    >0\}
\end{equation*}
is empty.
  \item The nonterminating probability is $p_{NT}(\rho)=\sum_{i\in I'}
  \<i|\rho|i\>.$
\end{enumerate}
\end{cor}

Now we are able to show that the quantum walk~(\ref{lp:qwalk}) in
Example~\ref{exam} is almost terminating. It is direct to calculate
that $P_X=\sum_{i\neq 1} |i\>\<i|\otimes I_2$ and
$$U_X=\sum_{i\neq 0,1}\left(|i\>\<i\oplus 1|\otimes |0\>\<+|+
|i\oplus 1\>\<i|\otimes |0\>\<-|\right).$$ With
Corollary~\ref{cor:normlessthan1} it suffices to prove that each
eigenvector of $U_X$ has its norm strictly less that 1. By
contradiction, suppose $U_X$ has an eigenvalue $\lambda$ with unit
norm, and one of the corresponding normalized eigenvector is
\begin{equation}\label{eq:psi}
|\psi\>=\sum_{i=0}^{n-1}(\alpha_i |i\>\otimes |+\>+\beta_i
|i\>\otimes |-\>),
\end{equation}
where $\sum_i (|\alpha_i|^2+|\beta_i|^2) =1$. Then we have
\begin{equation}\label{eq:lambdapsi}
\lambda|\psi\>=U_X|\psi\>=\sum_{i\neq 0,1}(\alpha_{i\oplus
1}|i\>\otimes |0\>+\beta_i|i\oplus 1\>\otimes |1\>).
\end{equation}
Comparing Eqs.~(\ref{eq:psi}) and (\ref{eq:lambdapsi}), we derive
further that
\begin{equation}\label{eq:ab+}
\alpha_i+\beta_i=\left\{
    \begin{array}{ll}
      0, & \mbox{if }\ i=0,1; \\
      \displaystyle \sqrt{2}\alpha_{i\oplus 1}/\lambda, &  \mbox{if }\
i\neq 0,1
    \end{array}
  \right.
\end{equation}
 and
\begin{equation}\label{eq:ab-}
\alpha_i-\beta_i=\left\{
    \begin{array}{ll}
      0, & \mbox{if }\ i=1,2; \\
      \displaystyle \sqrt{2}\beta_{i\ominus 1}/\lambda, &  \mbox{if }\
i\neq 1,2.
    \end{array}
  \right.
\end{equation}
On the other hand, since $|\lambda|=1$, we know from
Eq.~(\ref{eq:lambdapsi}) that
\begin{equation}
\sum_{i\neq 0,1} (|\alpha_{i\oplus
1}|^2+|\beta_i|^2)=\|\lambda|\psi\>\|^2=1.
\end{equation}
So we have:
\begin{equation}\label{eq:ab0}
\alpha_1=\alpha_2=\beta_0=\beta_1=0.
\end{equation}
Taking Eq.~(\ref{eq:ab0}) back into Eqs.~(\ref{eq:ab+}) and
(\ref{eq:ab-}) we can deduce that $\alpha_i=\beta_i=0$ for any $i$.
This is a contradiction.

To conclude this section, we observe some further properties of
almost terminating quantum loops. The following theorem indicates
that the notion of uniformly almost terminating loop coincides with
almost terminating loop.
\begin{thm}\label{thm:teralter} The quantum loop ~(\ref{eq: lp1}) is almost
terminating if and only if it is uniformly almost terminating.
\end{thm}
{\it Proof.} If loop ~(\ref{eq: lp1}) is almost terminating, then we
have $|\lambda_i|<1$ for any $i=1,\dots,l$ from Corollary
\ref{cor:normlessthan1}. Let $U_X=SJ(U_X)S^{-1}$ be the Jordan
decomposition of $U_X$. Then from Eq.~(\ref{eqn:pn}) we have:
\begin{equation*}
    p_{NT}^{n+}(\rho)= \|SJ(U_X)^{n-1}S^{-1}\rho_X^{1/2}\|^2.
\end{equation*}
By using the properties of matrix norm, we derive that
\begin{eqnarray*}
    p_{NT}^{n+}(\rho)&\leq& (\|S\|\cdot\|S^{-1}\|\cdot\|\rho_X^{1/2}\|)^2
    \|J(U_X)\|^{2(n-1)}\\
    &\leq& (\|S\|\cdot\|S^{-1}\|)^2
    \|J(U_X)\|^{2(n-1)}.
\end{eqnarray*}
Since $spec(J(U_X))\in [0,1)$, from Eq.~(\ref{eq: 8}) we can check
easily that $\|J(U_X)\|^{n}\ra 0$ when $n\ra \infty$. So for any
$\epsilon>0$, we can take $n(\epsilon)$ large enough such that
$$\|J(U_X)\|^{2(n(\epsilon)-1)}<\frac{\epsilon}{(\|S\|\cdot\|S^{-1}\|)^2}$$ Then we
have $p_{NT}^{n+}(\rho)<\epsilon$ for all $\rho$ whenever $n\geq
n(\epsilon)$. Thus loop ~(\ref{eq: lp1}) is uniformly almost
terminating. \hfill $\square$

\vspace{1em}

The next two theorems show that the notion of almost terminating
loop is sensitive. More explicitly, a small disturbance either on
the unitary transformation in the loop body or on the measurement in
the loop guard can make any quantum loop almost terminating.

We first need to prove a technical lemma.

\begin{lem}\label{lem:alsoeigen}
Suppose $|i\>$ is an eigenvector of $U_X$ and its corresponding
eigenvalue $\lambda_i$ has unit norm. Then:
\begin{enumerate}
\item $|i\>\in H_X$, and it is also an eigenvector of $U$ with an
eigenvalue of unit norm;
\item $P_{\overline{X}}U|i\>=0$.
\end{enumerate}
\end{lem}
{\it Proof.} Assume that $U_X|i\>=\lambda|i\>$ and $|\lambda|=1$.
First, we see that $\lambda P_X|i\>=P_XU_X|i\>=U_X|i\>=\lambda
|i\>.$ Thus, $P_X|i\>=|i\>$ and $|i\>\in H_X$. Furthermore, by the
Gram-Schmidt procedure we can find an orthonormal basis $\{|j\>\}$
for $H$ which contains $|i\>$. We assume that
$U|i\>=\sum_{j}\mu_j|j\>$ and $\sum_{j}|\mu_j|^{2}=1$. Then it holds
that $|\mu_i|=|\<i|U|i\>|=|\<i|P_XUP_X|i\>|=|\lambda|=1.$ This
implies that $\mu_j=0$ for all $j\neq i$, and $U|i\>=\mu_i|i\>$.
Finally, $P_{\overline{X}}U|i\>=\mu_i P_{\overline{X}}|i\>=0.$
\hfill $\square$

\begin{thm}\label{thm:disturbU} For any $M$, $X\neq spec(M)$ and $U$
in loop (\ref{eq: lp1}), and for any $\epsilon>0$, there exists a
unitary operator $U^{\prime}$ such that $\|U-U'\|<\epsilon$ and the
following loop is almost terminating:
\begin{equation*}\mathbf{while}\ (M\in X)\
\{\overline{q}:=U'\overline{q}\}.\end{equation*}
\end{thm}
{\it Proof.} By using Corollary~\ref{cor:normlessthan1}, we only
need to find a unitary operator $U'$ such that $\|U-U'\|<\epsilon$
and all eigenvalues of $P_XU'P_X$ have norms less than 1. On the
other hand, Lemma~\ref{lem:alsoeigen} implies that a necessary
condition for $P_XU'P_X$ to have an eigenvalue with unit norm is
that $U'$ has an eigenvector lying in the space $H_X$. Here $H_X$ is
the subspace with projector $P_X$. So we need only to show that we
can take $U'$ such that $\|U-U'\|<\epsilon$ and at the same time
none of the eigenvectors of $U'$ lies in $H_X$. To achieve this, we
first write out the spectrum decomposition of $U$ as $U=\sum_i \mu_i
|\psi_i\>\<\psi_i|$. If each $|\psi_i\>\not\in H_X$ then we have
done. Otherwise suppose $|\psi_{i_0}\>\in H_X$ for some $i_0$. From
$X\neq spec(M)$ there exists $j_0$ such that $|\psi_{j_0}\>\not\in
H_X$. Let
\begin{equation}
|\psi_{i_0}'\>=\sqrt{1-\delta}|\psi_{i_0}\>+\sqrt{\delta}|\psi_{j_0}\>,
\end{equation}
\begin{equation}
    |\psi_{j_0}'\>=\sqrt{1-\delta}|\psi_{j_0}\>-\sqrt{\delta}|\psi_{i_0}\>,
\end{equation}
and $|\psi_i'\>=|\psi_i\>$ for $i\neq i_0,j_0$. Here $\delta$ is a
very small but positive real number which will be determined later.
It is obvious that the set $|\psi_i'\>$ are also orthonormal, and
$|\psi_{i_0}'\>, |\psi_{j_0}'\>\not \in H_X$. Let
$U_1=\sum_{i}\mu_i|\psi_i'\>\<\psi_i'|$. Then the number of
eigenvectors of $U_1$ which lie in $H_X$ is strictly less than that
of $U$. Repeating the above steps we can finally find a sequence of
unitary matrices $U=U_0,U_1,\dots,U_d$, $d\leq K=dim(H)$, such that
all the eigenvectors of $U_d$ does not lie in $H_X$. Take $U'=U_d$
and notice that we can take $\delta$ small enough at each step such
that $\|U_i-U_{i+1}\|<\epsilon/K$. It then follows that
$\|U-U'\|\leq \sum_{i=0}^{d-1}\|U_i-U_{i+1}\| <\epsilon$. \hfill
$\square$

\begin{thm}\label{thm:disturbM} For any $M$, $X\neq spec(M)$ and $U$
in loop (\ref{eq: lp1}), and for any $\epsilon>0$, there exists an
observable $M^{\prime}$ with $spec(M')=spec(M)$ such that
$\|M'-M\|<\epsilon$ and the following loop is almost terminating:
\begin{equation*}\mathbf{while}\ (M'\in X)\
\{\overline{q}:=U\overline{q}\}\end{equation*}
\end{thm}
{\it Proof.} Similar to the proof of Theorem~\ref{thm:disturbU}, it
suffices to find $M'$ such that $spec(M')=spec(M)$ ,
$\|M-M'\|<\epsilon$, and none of the eigenvectors of $U$ lies in
$H_X'$, where $H_X'$ is the eigenspace of $M'$ with eigenvalues in
$X$. Let $\{|m_i\>\}_ {i=1}^K$ be an orthonormal basis of $H$ such
that $P_X=\sum_{i=1}^{k} |m_i\>\<m_i|$. Since $X\neq spec(M)$, we
have $1\leq k<K$.

Let $U=\sum_j \mu_j |\psi_j\>\<\psi_j|$ be the spectrum
decomposition of $U$. If all $|\psi_j\>\not\in H_X$ then we have
done. Otherwise assume $|\psi_{j_0}\>\in H_X$. Then there exists
$i_0\leq k$ such that $\<m_{i_0}|\psi_{j_0}\>\neq 0$. We take some
$k+1\leq i_1\leq K$ and put
\begin{equation}
|m_{i_0}^1\>=\sqrt{1-\delta}|m_{i_0}\>+\sqrt{\delta}|m_{i_1}\>,
\end{equation}
\begin{equation}
    |m_{i_1}^1\>=\sqrt{1-\delta}|m_{i_1}\>-\sqrt{\delta}|m_{i_0}\>,
\end{equation} where it is required that $0<\delta<1,$
and $|m_i^1\>=|m_i\>$ for $i\neq i_0 ,i_1$. It is easy to check that
the set $\{|m_i^1\>\}_{i=1}^{K}$ is also an orthonormal basis of
$H$. We write $P_{X}^1=\sum_{i=1}^k|m_i^1\>\<m_i^1|$. Let $H_X^1$ be
the subspace of $H$ with projector $P_X^1$. Then $|\psi_{j_0}\>\not
\in H_X^1$ because
$\<m_{i_1}^1|\psi_{j_0}\>=-\sqrt{\delta}\<m_{i_0}|\psi_{j_0}\>\neq
0.$ Furthermore, we can choose $\delta$ carefully such that the
eigenvectors of $U$ in $H_X^1$ is strictly less than that in $H_X$.
Indeed, if $|\psi_j\>\not\in H_X$ but $|\psi_j\>\in H_X^1$ then it
must hold that
\begin{equation}
    0=\<m_{i_1}^1|\psi_j\>=\sqrt{1-\delta}\<m_{i_1}|\psi_j\>-\sqrt{\delta}\<m_{i_0}|\psi_j\>.
\end{equation}
Thus, there are only finitely many $\delta$ which does not meet our
requirement.

Repeating the above steps we can finally find a sequence of
orthonormal bases $\{|m_i^l\>, i=1,\dots,K\}_{l=0}^d$ with $d\leq K$
such that $|m_i^0\>=|m_i\>$ for each $i\leq K$, and all the
eigenvectors of $U$ do not lie in $H_X^d$, the subspace of $H$ with
projector $P_X^d=\sum_{i=1}^k|m_i^d\>\<m_i^d|$. Let
$$P_m'=\sum_{P_m|m_i\>=|m_i\>}|m_i^d\>\<m_i^d|\ ,\  M'=\sum_{m\in
spec(M)} mP_m'.$$ Notice that we can take $\delta$ small enough at
each step such that
$$\max\{\|P_X^i-P_X^{i+1}\|,\|P_{\overline{X}}^i-P_{\overline{X}}^{i+1}\|\}
<\frac{\epsilon}{K\cdot\sum_m |m|}.$$ It then follows that
\begin{eqnarray*}
\|M-M'\|&\leq& \sum_m |m|\cdot\|P_m-P_m'\|\\
&\leq&  \sum_m |m| \left(\sum_{i=0}^{d-1}\|P_m^i-P_m^{i+1}\|\right)\\
&<&  \sum_m |m| \left(\sum_{i=0}^{d-1}\frac{\epsilon}{K\cdot\sum_m
|m|}\right) <\epsilon.
\end{eqnarray*} \hfill $\square$

\section{The Function Computed by a Quantum Loop}

In this section, we are going to give a representation of the
function computed by the loop~(\ref{eq: lp1}). First of all, we
consider the simple case that $U_X$ is normal.

\begin{thm}\label{prop:normalf}
Suppose $U_X$ is normal and its spectrum decomposition is given by
Eq.~(\ref{eq:sdecom}). Then the function computed by loop~(\ref{eq:
lp1}) is as follows:
\begin{equation*}
  F(\rho) = \displaystyle P_{\overline{X}}\rho
P_{\overline{X}} + \sum_{i,j\in
I}\frac{\<i|\rho|j\>}{1-\lambda_i\lambda_j^*}P_{\overline{X}}U|i\>\<j|
  U^\dag P_{\overline{X}}
\end{equation*}
where $I \equiv \{i\ |\ \ 1\leq i\leq k\ ,\ |\lambda_i|<1\}$.
\end{thm}
{\it Proof.} For any $n\geq 0$, we have from Eq.~(\ref{eq:sdecom})
that
$$P_{\overline{X}}U U_X^n=\sum_{i=1}^k \lambda_i^n
P_{\overline{X}}U|i\>\<i|=\sum_{i\in I} \lambda_i^n
P_{\overline{X}}U|i\>\<i|$$ where the second equality is due to
Lemma~\ref{lem:alsoeigen}.2. Taking this equation into
Eq.~(\ref{eqn:frho}) we have:
\begin{equation*}
F(\rho)  = \displaystyle P_{\overline{X}}\rho P_{\overline{X}} +
\sum_{i,j\in I}\left(\sum_{n=0}^{\infty} \lambda_i^n
  \lambda_j^{*n}\right)\<i|\rho_X|j\>P_{\overline{X}}U|i\>\<j| U^\dag
  P_{\overline{X}}.
\end{equation*}
Then the result follows by using Lemma \ref{lem:alsoeigen}.1. \hfill
$\square$

\vspace{1em}

We now turn to consider the general case where $U_X$ is not
necessary to be normal. To this end, the following lemmas are
needed.

\begin{lem}(Schur's unitary triangularization; \cite{Bh91}) Given $(k\times k)$-complex matrix $A$
with eigenvalues $\lambda_1,\dots,\lambda_k$ in any prescribed
order, there exists a $(k\times k)$ unitary matrix $V$ such that
$A=VTV^\dag$, where $T$ is upper triangular with diagonal entries
$T_{ii}=\lambda_i$, $i=1,\dots,k$.
\end{lem}

\begin{lem}\label{lem:equal0} Let $U_X=VTV^\dag$ be the Schur's triangularization
of $U_X$. Then for any $1\leq i\leq k$, if $|T_{ii}|=1$ then
$T_{ij}=T_{ji}=0$ for all $j\neq i$.
\end{lem}
{\it Proof.} To prove this lemma, we need only to notice $T^\dag
T=V^\dag U_X^\dag U_X V\leq I$, and so for any $i$, the Euclidean
norms of the $i$-th row and the $i$-th column of $T$ must be less
than or equal to $1$. \hfill $\square$

\begin{lem}\label{lem:size1} For each Jordan block $J_r(\lambda)$ in the Jordan normal form
of $U_X$, if $|\lambda|=1$, then $r=1$. That is, each Jordan block
corresponding to unit norm eigenvalues has size 1.
\end{lem}
{\it Proof.} Suppose $U_X=VTV^\dag$ is the Schur's triangularization
of $U_X$, and the diagonal entries of $T$ have been arranged in
decreasing order of their norms, i.e.,
$1=|T_{11}|=\dots=|T_{tt}|>|T_{t+1,t+1}|\geq\dots\geq |T_{kk}|$ for
some $t$. Then from Lemma~\ref{lem:equal0}, $T$ must have the form
$$T=\left(
      \begin{array}{ccc|c}
        T_{11} &  &   &\\
         & \ddots &   & \\
         &  & T_{tt}  &\\
         \hline
         &  &  &  T' \\
      \end{array}
    \right)
$$
where $T'$ is $(k-t)\times (k-t)-$dimensional and none of its
eigenvalues has unit norm. Let $T'=S'J(T')S'^{-1}$ be the Jordan
decomposition of $T'$, and let
$$S=V\left(
      \begin{array}{c|c}
      I_t  & \\
      \hline
         &  S'
      \end{array}
    \right), J=\left(
      \begin{array}{ccc|c}
        T_{11} &  &  &\\
         & \ddots &   &\\
         &  & T_{tt} &\\
         \hline
         &  &  & J(T') \\
      \end{array}
    \right)$$
It is easy to check that $SJS^{-1}$ is the Jordan decomposition of
$U_X$. Then the result holds from the uniqueness of Jordan normal
form in the sense presented in Lemma~\ref{lem: 6.2}. \hfill
$\square$

\begin{lem}\label{lem: jrn} Let $J_r(\lambda)$ be a $(r\times
r)-$Jordan block, $|\lambda|<1$, and
$\mathbf{v}=(v_1,v_2,\dots,v_r)^{T}\in \mathbf{C}^{r}$. Then
\begin{equation}\label{eq:limjrn}
\begin{array}{rcl}
\displaystyle\sum_{n=0}^\infty
J_r(\lambda)^{n}\mathbf{v}&=&\displaystyle(\sum_{i=0}^{r-1}
f_i(\lambda)v_{i+1}, \sum_{i=0}^{r-2}f_i(\lambda)v_{i+2}, \dots,\\
\\&&f_0(\lambda)v_{r-1}+f_1(\lambda)v_r,f_0(\lambda)v_r)^{T},
\end{array}
\end{equation}
where $$f_i(x)=\frac{\displaystyle d^i (1-x)^{-1}}{\displaystyle
i!\ dx^i}.$$
\end{lem}
{\it Proof.} For any $1\leq m\leq r$, we can see from Eqs.~(\ref{eq:
8}) and (\ref{eq: 9}) that the $m-$component of vector
$\sum_{n=0}^\infty J_r(\lambda)^{n}\mathbf{v}$ is
\begin{eqnarray*}
\left(\sum_{n=0}^\infty J_r(\lambda)^{n}\mathbf{v}\right)_m &=&
\sum_{i=0}^{r-m}  \sum_{n=0}^\infty \left(\begin{array}{cc}n\\
i\end{array}\right)\lambda^{n-i}v_{i+m}\\
&=& \sum_{i=0}^{r-m} \left(\left. \frac{d^i \sum_{n=0}^\infty
x^n}{\displaystyle i!\
dx^i}\right|_{x=\lambda}\right)  v_{i+m}\\
&=& \sum_{i=0}^{r-m} f_i(\lambda)  v_{i+m}.
\end{eqnarray*}
The convergence of the above series is guaranteed by the assumption
that $|\lambda|<1$. \hfill $\square$

\vspace{1em}

Now we are able to present the main result of this section.

\begin{thm}\label{prop: 6.4} Suppose that $S$, $J(U_X)$, $J_{k_i}(\lambda_i)$ and
$\mathbf{v}_i$ $(1\leq i\leq l)$ are given as in Theorem~\ref{prop:
5.1}. Without loss of generality, we assume that the Jordan blocks
of $J(U_X)$ have been arranged in the decreasing order of
$|\lambda_i|$, i.e. $1=|\lambda_1|= \dots
=|\lambda_t|>|\lambda_{t+1}|\geq\dots\geq|\lambda_l|$. Then the
output $F(|\psi\>)$ of the loop~(\ref{eq: lp1}) with input $|\psi\>$
is a $(K-k)$-dimensional vector lying in the subspace
$H_{\overline{X}}$:
\begin{equation}
    F(|\psi\>)=(|\psi\> + US\mathbf{u})_{\overline{X}},
\end{equation}
where
$\mathbf{u}=(\mathbf{0},\mathbf{u}_{t+1}^T,\dots,\mathbf{u}_{l}^T,\mathbf{0})^T$
is a $K$-dimensional vector. Here the former and the latter zero
vectors have dimensions $t$ and $K-k$, respectively, and for $i=
t+1,\dots,l$, $\mathbf{u}_i=\sum_{n=0}^\infty
J_{k_i}(\lambda_i)^{n}\mathbf{v_i}$ is given in
Eq.~(\ref{eq:limjrn}).
\end{thm}
{\it Proof.} Under the assumption of the theorem, we have
$k_1=\dots=k_t=1$ by using Lemma~\ref{lem:size1}. Then for any $i=
1,\dots,t$,
$$U_XS|m_i\>=SJ(U_X)|m_i\>=\lambda_i S|m_i\>,$$
or in other words, $S|m_i\>$ is an eigenvector of $U_X$ with its
corresponding eigenvalue having unit norm. So we have
$P_{\overline{X}}US|m_i\>=0$ from Lemma \ref{lem:alsoeigen} 2.

On the other hand, from Eq.~(\ref{eqn:frho}) we have
\begin{eqnarray}
F(|\psi\>) &=& P_{\overline{X}}|\psi\> +
\displaystyle\sum_{n=0}^{\infty} P_{\overline{X}}U U_X^n|\psi\>_X
\nonumber
\\
&=&P_{\overline{X}}|\psi\> + \displaystyle\sum_{n=0}^{\infty}
P_{\overline{X}}USJ(U_X)^nS^{-1}|\psi\>_X\nonumber\\
&=&P_{\overline{X}}|\psi\> +
P_{\overline{X}}US\displaystyle\sum_{n=0}^{\infty} J(U_X)^n
\mathbf{v}'.\label{eq:fpsi}
\end{eqnarray} Here $\mathbf{v}'=(\mathbf{0},\mathbf{v}_{t+1}^T,\dots,\mathbf{v}_l^T)^T$ and the zero vector
$\mathbf{0}$ has dimension $t$. Then the result holds by using
Lemma~\ref{lem: jrn} and rewriting Eq.~(\ref{eq:fpsi}) into vector
form. \hfill $\square$

\vspace{1em}

Although we only consider pure input states in
Theorems~\ref{prop:normalf} and \ref{prop: 6.4}, they may be used to
calculate the outputted state $F(\rho)$ of loop~(\ref{eq: lp1}) for
any mixed input state $\rho$ by noting that $F(\rho)=\sum_i p_i
F(|\psi_i\>)$, where $\rho=\sum_i p_i |\psi_i\>\<\psi_i|$ is the
spectrum decomposition of $\rho$.

\section{Some Illustrative Examples}

To illustrate further the notions introduced and the results
obtained in this paper, we consider two simplest classes of quantum
loops.

\subsection{Single qubit loops}

Let $M$ be an observable in the $2-$dimensional Hilbert space $H_2$.
Then we have $M=m_1|m_1\rangle \langle m_1|+m_2|m_2\rangle \langle
m_2|,$ where $m_1,m_2$ are the eigenvalues of $M$, and $|m_i\rangle$
is the eigenvector of $M$ corresponding to $m_i$ $(i=1,2)$. A single
qubit loop can be written as follows:
\begin{equation}\label{eq: lp2}\mathbf{while}\ (M=m_i)\ \{q:=Uq\},\end{equation} where
$U$ is a unitary operation on a single qubit, and $i=1,2$. Without
any loss of generality we may assume that $m_1\neq m_2$ and $i=1$.

Note that the function $F$ defined by the loop~(\ref{eq: lp2}) is
from $\mathcal{D}(H_2)$ to $\mathcal{D}^{-}(span\{|m_2\rangle\})$.
Since $span\{|m_2\rangle\}$ is the one-dimensional Hilbert space,
$\mathcal{D}^{-}(span\{|m_2\rangle\})$ can be identified with the
unit interval $[0,1]$. Thus, the function $F$ computed by the
loop~(\ref{eq: lp2}) is a mapping from $\mathcal{D}(H_2)$ into
$[0,1]$. A simple application of Theorems~\ref{prop: 5.1}, \ref{thm:
2} and~\ref{prop: 6.4} leads to the following:

\begin{lem}\label{lem: 4.1} Let $\rho\in \mathcal{D}(H_2)$ be the input
state to the single qubit loop program~(\ref{eq: lp2}). Then:
\begin{enumerate}
\item if $\langle m_1|\rho|m_1\rangle =0$ or $\langle
m_1|U|m_1\rangle =0$, then the loop~(\ref{eq: lp2}) terminates, and
$F(\rho)=1$;

\item if $|\langle m_1|U|m_1\rangle|<1$, then the loop~(\ref{eq: lp2}) almost terminates, and $F(\rho)=1$;

\item if $\langle m_1|\rho|m_1\rangle >0$ and $|\langle
m_1|U|m_1\rangle |=1$, then the loop~(\ref{eq: lp2}) does not
terminate, and $F(\rho)=\langle
m_2|\rho|m_2\rangle$.\end{enumerate}\end{lem}

Now we further consider the special case that the input is a pure
state. To this end, we shall need the following:

\begin{lem}\label{lem: 4.2} (\cite{BBC95, NC00}) Each unitary operation on a
single qubit can be written in the form of
$U=e^{i\alpha}R_z(\beta)R_y(\gamma)R_z(\delta),$ where $\alpha$,
$\beta$, $\gamma$ and $\delta$ are real numbers,
$$R_y(\theta)=\left(\begin{array}{cc}\cos\frac{\theta}{2} &
-\sin\frac{\theta}{2}\\\sin\frac{\theta}{2} &
\cos\frac{\theta}{2}\end{array}\right)$$ and
$$R_z(\theta)=\left(\begin{array}{cc}e^{-\frac{i\theta}{2}} & 0\\
0 & e^{\frac{i\theta}{2}}\end{array}\right)$$ are the rotation
operators about $y$ and $z$ axes, respectively.\end{lem}

To simplify the presentation, we further suppose that the
measurement is performed on the computational basis. Combining
Lemmas~\ref{lem: 4.1} and~\ref{lem: 4.2} we obtain:

\begin{prop}\label{prop: 4.1} Suppose that $|\psi\rangle=a_0|0\rangle
+a_1|1\rangle$ is the input to the single qubit loop program:
\begin{equation}\label{eq: lp3}\mathbf{while}\ (q=0)\ \{q:=Uq\},\end{equation} where the loop condition
$(q=0)$ means that the outcome of a measurement on the computational
basis $|0\rangle, |1\rangle$ is $0$, and the unitary operator $U$ is
given as in Lemma \ref{lem: 4.2}. Then
\begin{enumerate}\item
 if $a_0=0$ or $\gamma =(2n+1)\pi$ for some integer $n$, then the
loop~(\ref{eq: lp3}) terminates;
\item if $a_0\neq 0$ and $\gamma =2n\pi$ for some integer $n$, then the
loop~(\ref{eq: lp3}) does not terminate;
\item if $\gamma \neq n\pi$ for any integer $n$, then the loop~(\ref{eq: lp3})
almost terminates.
\end{enumerate}
A similar conclusion holds if the guard condition $(q=0)$ in the
loop~(\ref{eq: lp3}) is replaced by $(q=1)$, which means that the
result $1$ occurs when performing a measurement on the computational
basis $|0\rangle, |1\rangle$.\end{prop}

It is interesting to note from the above proposition that
termination of the loop~(\ref{eq: lp3}) depends only upon the
parameter $\gamma$, and it is irrelevant to the other parameters
$\alpha, \beta$ and $\delta$. Moreover, we see that the
loop~(\ref{eq: lp3}) is terminating if $\gamma =(2n+1)\pi$ for some
integer $n$, and it is (uniformly) almost terminating if $\gamma\neq
n\pi$ for any integer $n$.

From Lemma \ref{lem: 4.1}, it is easy to see that in Proposition
~\ref{prop: 4.1} for the cases 1 and 2, we have $F(|\psi\rangle)=1$,
and for the case 3, $F(|\psi\rangle)=|a_1|^{2}$.

The most frequently used single qubit gates are the four Pauli
matrices:
$$I=\left(\begin{array}{cc}1 & 0\\ 0 & 1\end{array}\right),\ X=\left(\begin{array}{cc}0 & 1\\ 1 & 0\end{array}\right),$$
$$Y=\left(\begin{array}{cc}0 & -i\\ i & 0\end{array}\right),\ Z=\left(\begin{array}{cc}1 & 0\\ 0 & -1\end{array}\right),$$ the Hadamard
gate:
$$H=\frac{1}{\sqrt{2}}\left(\begin{array}{cc}1 & 1\\
1 & -1\end{array}\right),$$ the phase gate:
$$S=\left(\begin{array}{cc}1 & 0\\ 0 & i\end{array}\right),$$ and the $\frac{\pi}{8}$
gate:
$$T=\left(\begin{array}{cc}1 & 0\\
0 & exp(\frac{i\pi}{4})\end{array}\right).$$ Applying Proposition
~\ref{prop: 4.1} to these gates, we obtain:

\begin{cor} For a single qubit input state $|\psi\rangle
=a_0 |0\rangle + a_1 |1\rangle$, the loop~(\ref{eq: lp3}) always
terminates when $U=X$ or $Y$, it almost terminates when $U=H$, and
it does not terminate when $U=I,Z,S$ or $T$ provided $a_0\neq
0$.\end{cor}

\subsection{Two qubit loops defined by controlled operations}

As the second example we consider a typical class of two qubit
gates, namely, controlled operations. Suppose that $U$ is a single
qubit unitary operation. Then the controlled-$U$ gate is defined by
the following $4\times 4$ matrix:
$$C(U)=\left(\begin{array}{cc}I & 0\\ 0 & U\end{array}\right),$$ where $I$ is
the $2\times 2$ unit matrix. For a two qubit system, the measurement
$M$ on the computational basis $|00\rangle, |01\rangle, |10\rangle$
and $|11\rangle$ has four possible outcomes $00$, $01$, $10$ and
$11$, where we use $i_1i_2$ to indicate the measurement result
$q_1=i_1$ and $q_2=i_2$ for any $i_1,i_2\in \{0,1\}$.  Thus, the two
qubit quantum loop defined by controlled operation $C(U)$ may be
written as follows:
\begin{equation}\label{eq: lp4}\mathbf{while}\ (M\in X)\
\{q_1,q_2:=C(U)q_1,q_2\},\end{equation} where $X\subseteq \{00, 01,
10, 11\}$.

The following proposition carefully examines the behavior of this
loop for various choices of $X$ except the trivial cases
$X=\emptyset$ or $X=\{00, 01, 10, 11\}$.

\begin{prop} Let pure state $|\psi\rangle
=a_{00}|00\rangle+a_{01}|01\rangle+a_{10}|10\rangle+a_{11}|11\rangle$
be the input of the loop program~(\ref{eq: lp4}). Suppose that
$U=(U_{ij})_{i,j=0}^{1}$ is the matrix representation of $U$
according to the basis $\{|0\rangle, |1\rangle\}$, that is,
$U_{ij}=\langle i|U|j\rangle$ for any $i,j\in \{0,1\}$.
\begin{enumerate}

\item If $X=\{00\}$, then $p_{NT}=|a_{00}|^{2}$,
$F(|\psi\rangle)=a_{01}|01\rangle+a_{10}|10\rangle+a_{11}|11\rangle$,
the loop~(\ref{eq: lp4}) terminates if $a_{00}=0$, and it does not
terminates if $a_{00}\neq 0$.

\item If $X=\{01\}$, then $p_{NT}=|a_{01}|^{2}$,
$F(|\psi\rangle)=a_{00}|00\rangle+a_{10}|10\rangle+a_{11}|11\rangle$,
the loop~(\ref{eq: lp4}) terminates if $a_{01}=0$, and it does not
terminates if $a_{01}\neq 0$.

\item Let $X=\{10\}$. If $a_{10}=0$ or $U_{00}=0$, then the loop~(\ref{eq: lp4})
terminates. If $a_{10}=0$ or $|U_{00}|<1$, then it almost
terminates, and \begin{equation*}\begin{split}F(|\psi\rangle) &
=\left(\begin{array}{ccc}|a_{00}|^{2} & a_{00}a_{01}^{\ast} &
a_{00}a_{11}^{\ast}\\
a_{01}a_{00}^{\ast} & |a_{01}|^{2} & a_{01}a_{11}^{\ast}\\
a_{11}a_{00}^{\ast} & a_{11}a_{01}^{\ast}&
|a_{10}|^{2}+|a_{11}|^{2}\end{array}\right)\\ & \in
\mathcal{D}(span\{|00\rangle, |01\rangle,
|11\rangle\}).\end{split}\end{equation*} If $a_{10}\neq 0$ and
$U_{00}=1$, then it does not terminate, and
$F(|\psi\rangle)=a_{00}|00\rangle+a_{01}|01\rangle+a_{11}|11\rangle$.

\item Let $X=\{11\}$. If $a_{11}=0$ or $U_{11}=0$, then the loop~(\ref{eq: lp4})
terminates. If $a_{11}=0$ or $|U_{11}|<1$, then it almost
terminates, and \begin{equation*}\begin{split}F(|\psi\rangle) &
=\left(\begin{array}{cccc}|a_{00}|^{2} & a_{00}a_{01}^{\ast} &
a_{00}a_{10}^{\ast}\\
a_{01}a_{00}^{\ast} & |a_{01}|^{2} & a_{01}a_{10}^{\ast} \\
a_{10}a_{00}^{\ast} & a_{10}a_{01}^{\ast} &
|a_{10}|^{2}+|a_{11}|^{2}
\end{array}\right)\\ & \in \mathcal{D}(span\{|00\rangle, |01\rangle, |10\rangle\}).
\end{split}\end{equation*} If $a_{11}\neq 0$ and
$U_{11}=1$, then it does not terminate, and
$F(|\psi\rangle)=a_{00}|00\rangle+a_{01}|01\rangle+a_{10}|10\rangle$.

\item If $X=\{00,01\}$, then
$p_{NT}=|a_{00}|^{2}+|a_{01}|^{2}$, $F(|\psi\rangle)=
a_{10}|10\rangle+a_{11}|11\rangle,$ the loop~(\ref{eq: lp4})
terminates if $a_{00}=a_{01}=0$, and it does not terminate if
$a_{00}\neq 0$ or $a_{01}\neq 0$.

\item If $X=\{10,11\}$, then
$p_{NT}=|a_{10}|^{2}+|a_{11}|^{2}$, $F(|\psi\rangle)=
a_{00}|00\rangle+a_{01}|01\rangle$, the loop~(\ref{eq: lp4})
terminates if $a_{10}=a_{11}=0$, and it does not terminate if
$a_{10}\neq 0$ or $a_{11}\neq 0$.

\item Let $X=\{00, 10\}$. Then we have:
$$p_{NT}=\begin{cases}
|a_{00}|^{2}, & \mbox{if } |U_{00}|<1,\\
|a_{00}|^{2}+|a_{10}|^{2}, & \mbox{if } |U_{00}|=1.
\end{cases}$$ $F(|\psi\rangle)\in \mathcal{D}^{-}(span\{|01\rangle, |11\rangle\})$ is given as follows: for
the case of $|U_{00}|=1$, $F(|\psi\rangle) =a_{01}|01\rangle +
a_{11}|11\rangle$, and for the case of $|U_{00}|<1$,
\begin{equation*}F(|\psi\rangle) =\left(\begin{array}{cc}|a_{01}|^{2} & a_{01}a_{11}^{\ast}
\\ a_{11}a_{01}^{\ast} & |a_{10}|^{2}+|a_{11}|^{2}\end{array}\right).\end{equation*} If $a_{00}=0$, and
$a_{10}=0$ or $U_{00}=0$, then the loop~(\ref{eq: lp4}) terminates,
if $a_{00}=0$, and $a_{10}=0$ or $|U_{00}|<1$, then it almost
terminates, and if $a_{00}\neq 0$, or $a_{10}\neq 0$ and
$|U_{00}|=1$, then it does not terminate.

\item Let $X=\{00, 11\}$. Then we have:
$$p_{NT}=\begin{cases}|a_{00}|^{2}, & \mbox{if } |U_{11}|<1,\\
|a_{00}|^{2}+|a_{11}|^{2}, & \mbox{if } |U_{11}|=1.\end{cases}$$
$F(|\psi\rangle)\in \mathcal{D}^{-}(span\{|01\rangle, |10\rangle\})$
is given as follows: for the case of $|U_{11}|=1$, $F(|\psi\rangle)
=a_{01}|01\rangle + a_{10}|10\rangle$, and for the case of
$|U_{11}|<1$,
\begin{equation*}F(|\psi\rangle) =\left(\begin{array}{cc}|a_{01}|^{2} & a_{01}a_{10}^{\ast}
\\ a_{10}a_{01}^{\ast} & |a_{10}|^{2}+|a_{11}|^{2}\end{array}\right).\end{equation*}
If $a_{00}=0$, and $a_{11}=0$ or $U_{11}=0$, the the loop~(\ref{eq:
lp4}) terminates, if $a_{00}=0$ and $|U_{11}|<1$, or $a_{00}=0$ and
$a_{11}=0$, then it almost terminates, and if $a_{00}\neq 0,$ or
$a_{11}\neq 0$ and $|U_{11}|=1$, then it does not terminate.

\item Let $X=\{01, 10\}$. Then we have:
$$p_{NT}=\begin{cases}|a_{01}|^{2}, & \mbox{if } |U_{00}|<1,\\
|a_{01}|^{2}+|a_{10}|^{2}, & \mbox{if } |U_{00}|=1.\end{cases}$$
$F(|\psi\rangle)\in \mathcal{D}^{-}(span\{|00\rangle, |11\rangle\})$
is given as follows: for the case of $|U_{00}|=1$, $F(|\psi\rangle)
=a_{00}|00\rangle + a_{11}|11\rangle$, and for the case of
$|U_{00}|<1$,
\begin{equation*}F(|\psi\rangle) =\left(\begin{array}{cc}|a_{00}|^{2} & a_{00}a_{11}^{\ast}
\\ a_{11}a_{00}^{\ast} & |a_{10}|^{2}+|a_{11}|^{2}\end{array}\right).\end{equation*}
If $a_{01}=0$, and $a_{10}=0$ or $U_{00}=0$, the the loop~(\ref{eq:
lp4}) terminates, if $a_{01}=0$, and $|U_{00}|<1$ or $a_{10}=0$,
then it almost terminates, and if $a_{01}\neq 0,$ or $a_{10}\neq 0$
and $|U_{00}|=1$, then it does not terminate.

\item Let $X=\{01, 11\}$. Then we have:
$$p_{NT}=\begin{cases}|a_{01}|^{2}, & \mbox{if } |U_{11}|<1,\\
|a_{01}|^{2}+|a_{11}|^{2}, & \mbox{if } |U_{11}|=1.\end{cases}$$
$F(|\psi\rangle)\in \mathcal{D}^{-}(span\{|00\rangle, |10\rangle\})$
is given as follows: for the case of $|U_{11}|=1$, $F(|\psi\rangle)
=a_{00}|00\rangle + a_{10}|10\rangle$, and for the case of
$|U_{11}|<1$,
\begin{equation*}F(|\psi\rangle) =\left(\begin{array}{cc}|a_{00}|^{2} & a_{00}a_{10}^{\ast}
\\ a_{10}a_{00}^{\ast} & |a_{10}|^{2}+|a_{11}|^{2}\end{array}\right).\end{equation*}
If $a_{01}=0$, and $a_{11}=0$ or $U_{11}=0$, the the loop~(\ref{eq:
lp4}) terminates, if $a_{01}=0$, and $|U_{11}|<1$ or $a_{11}=0$,
then it almost terminates, and if $a_{01}\neq 0,$ or $a_{11}\neq 0$
and $|U_{11}|=1$, then it does not terminate.

\item Let $X=\{00, 01, 10\}$. Then we have: $$p_{NT}=\begin{cases}|a_{00}|^{2}+|a_{01}|^{2},
& \mbox{if } |U_{00}|<1,\\ |a_{00}|^{2}+|a_{01}|^{2}+|a_{10}|^{2}, &
\mbox{if } |U_{00}|=1,\end{cases}$$ and $F(|\psi\rangle)\in
\mathcal{D}^{-}(span \{|11\rangle\})\cong [0,1]$ is given by
$$F(|\psi\rangle)=\begin{cases}|a_{10}|^{2}+|a_{11}|^{2}, & \mbox{if
}|U_{00}|<1,\\ |a_{11}|^{2}, & \mbox{if }|U_{00}|=1.\end{cases}$$ If
$a_{00}=a_{01}=0$, and $a_{10}=0$ or $U_{00}=0$, then the
loop~(\ref{eq: lp4}) terminates, if $a_{00}=a_{01}=0$, and
$|U_{00}|<1$ or $a_{10}=0$, then it almost terminates, and if
$a_{00}\neq 0$, or $a_{01}\neq 0$, or $a_{10}\neq 0$ and
$|U_{00}|=1$, then it does not terminate.

\item Let $X=\{00, 01, 11\}$. Then we have: $$p_{NT}=\begin{cases}|a_{00}|^{2}+|a_{01}|^{2},
& \mbox{if } |U_{11}|<1,\\ |a_{00}|^{2}+|a_{01}|^{2}+|a_{11}|^{2}, &
\mbox{if } |U_{11}|=1,\end{cases}$$ and $F(|\psi\rangle)\in
\mathcal{D}^{-}(span \{|10\rangle\})\cong [0,1]$ is given by
$$F(|\psi\rangle)=\begin{cases}|a_{10}|^{2}+|a_{11}|^{2}, & \mbox{if
}|U_{11}|<1,\\ |a_{10}|^{2}, & \mbox{if }|U_{11}|=1.\end{cases}$$ If
$a_{00}=a_{01}=0$, and $a_{11}=0$ or $U_{11}=0$, then the
loop~(\ref{eq: lp4}) terminates, if $a_{00}=a_{01}=0$, and
$|U_{11}|<1$ or $a_{11}=0$, then it almost terminates, and if
$a_{00}\neq 0$, or $a_{01}\neq 0$, or $a_{11}\neq 0$ and
$|U_{11}|=1$, then it does not terminate.

\item Let $X=\{00, 10, 11\}$. Then
$p_{NT}=|a_{00}|^{2}+|a_{10}|^{2}+|a_{11}|^{2}$ and
$F(|\psi\rangle)=|a_{01}|^{2}\in \mathcal{D}^{-}(span
\{|01\rangle\})\cong [0,1]$. If $a_{00}=a_{10}=a_{11}=0$, then the
loop~(\ref{eq: lp4}) terminates, otherwise it does not terminate.

\item Let $X=\{01, 10, 11\}$. Then
$p_{NT}=|a_{01}|^{2}+|a_{10}|^{2}+|a_{11}|^{2}$ and
$F(|\psi\rangle)=|a_{00}|^{2}\in \mathcal{D}^{-}(span
\{|00\rangle\})\cong [0,1]$. If $a_{01}=a_{10}=a_{11}=0$, then the
loop~(\ref{eq: lp4}) terminates, otherwise it does not terminate.
\end{enumerate}\end{prop}

Note that termination of the loop~(\ref{eq: lp4}) is irrelevant to
the unitary operator $U$, and it only depends on the input state
$|\psi\rangle$ when $X=\{00\}, \{01\}, \{00,01\}, \{10,11\},
\{00,10,11\}$ or $\{01, 10, 11\}$. For the other cases, termination
of the loop defined by the CNOT gate is summarized in the following:

\begin{cor}\label{cor: 4.1}Suppose that $C(U)$ is the CNOT gate $C(X)$, where $X=NOT$ is the second Pauli gate.
\begin{enumerate}
\item Let $X=\{10\}$ or $\{11\}$. Then the loop~(\ref{eq: lp4}) always
terminates.
\item Let $X=\{00, 10\}$ or $\{00, 11\}$. Then the loop~(\ref{eq:
lp4}) terminates if $a_{00}=0$, otherwise it does not terminate.
\item Let $X=\{01, 10\}$ or $\{01, 11\}$. Then the loop~(\ref{eq:
lp4}) terminates if $a_{01}=0$, otherwise it does not terminate.
\item Let $X=\{00, 01, 10\}$ or $\{00, 01, 11\}$. Then the loop~(\ref{eq:
lp4}) terminates if $a_{00}=a_{01}=0$, otherwise it does not
terminate.
\end{enumerate}
\end{cor}

\section{Conclusion}

Exploitation of the full power of loop construct in quantum
computation requires a deep understanding of the computational
mechanism of quantum loop programs. In this paper, we introduced a
general scheme of quantum loop programs, the behaviors of quantum
loops are carefully analyzed, including termination, almost
termination, and sensitivity, and a matrix-summation representation
of the function computed by a quantum loop is found.

This paper is merely an initial step toward a thorough understanding
of quantum loop programs, and many important problems remain open.
First, the bodies of quantum loops that we considered are unitary
transformations. If a quantum loop is allowed to be embedded into
another quantum loop, then as was observed in Section 3, the body of
the latter loop is not a unitary operator but a super-operator in
general. Therefore, it is an interesting topic for further studies
to find conditions for termination and almost termination of quantum
loops in which the loop bodies are super-operators. Second, we
demonstrated the expressive power of quantum loops by presenting a
loop description of quantum walks. It would be very interesting to
find more computational problems that cannot be expressed or solved
without quantum loops. In general, the study of loop programs is a
very important area of computer programming methodology.
Reconsideration of some fundamental problems from this area, say
loop invariants and proof rules, in the quantum setting is a great
challenge.

\section*{Acknowledgement}

The authors would like to thank Mr. Zhengfeng Ji who suggested to
describe quantum walks as quantum loop programs. The first draft of
this paper was prepared when the first author was visiting Center of
Logic and Computation, Department of Mathematics, Instituto Superior
T\'{e}cnico, Technical University of Lisboa, Portugal. He is very
grateful to Professors Am\'{i}lcar Sernadas, Cristina Sernadas and
Paulo Mateus for their stimulating discussions and for providing the
excellent working environment.

\end{document}